\documentclass{ifacconf}

\usepackage{natbib}  

\usepackage{graphicx}
\usepackage{caption}

\usepackage{multicol}
\usepackage{amsmath,amssymb}
\usepackage{amsfonts}
\usepackage{textcomp}
\usepackage{psfrag}
\usepackage{multimedia}
\usepackage{fancybox}

\newcommand{\vect}[1]{\ensuremath{\boldsymbol{\mathrm{#1}}}}

\newtheorem{Proposition}{Proposition}
\newtheorem{Corollary}{Corollary}
\newtheorem{Lemma}{Lemma}

\newtheorem{Remark}{Remark}

\usepackage[usenames,dvipsnames]{color}
\definecolor{wheat}{rgb}{0.96,0.87,0.70}
\definecolor{mario}{rgb}{0.8,0.8,1}
\definecolor{seb}{rgb}{0.8,1,0.8}

\newcommand {\matr}[2]{\left[\begin{array}{#1}#2\end{array}\right]}

\newcounter{lastnote}

\begin{document} 
\begin{frontmatter}

\title{Safe Reinforcement Learning via Projection on a Safe Set: How to Achieve Optimality?} 

\author[First]{Sebastien Gros} 
\author[Second]{Mario Zanon} 
\author[Second]{Alberto Bemporad} 

\address[First]{Norwegian University of Technology, NTNU}
\address[Second]{IMT School for Advanced Studies Lucca}




%
%

\begin{abstract} For all its successes, Reinforcement Learning (RL) still struggles to deliver formal guarantees on the closed-loop behavior of the learned policy. Among other things, guaranteeing the safety of RL with respect to safety-critical systems is a very active research topic. Some recent contributions propose to rely on projections of the inputs delivered by the learned policy into a safe set, ensuring that the system safety is never jeopardized. Unfortunately, it is unclear whether this operation can be performed without disrupting the learning process. This paper addresses this issue. The problem is analysed in the context of $Q$-learning and policy gradient techniques. We show that the projection approach is generally disruptive in the context of $Q$-learning though a simple alternative solves the issue, while simple corrections can be used in the context of policy gradient methods in order to ensure that the policy gradients are unbiased. The proposed results extend to safe projections based on robust MPC techniques.
\end{abstract}

\end{frontmatter}

\section{Introduction}
Reinforcement Learning (RL) is a tool for tackling optimal control from data. RL methods seek to increase the closed-loop performance of the control policy deployed on the system as observations are collected. RL methods often rely on Deep Neural Networks (DNN) to carry the policy approximation $\vect\pi_{\vect\theta}$. Control policies based on DNNs provide limited opportunities for formal verifications of the resulting closed-loop behavior, and for imposing hard constraints on the evolution of the state of the real system. The development of safe RL methods is currently an open field of research \citep{Garcia2015}. 

In order to tackle safety issues in RL, it has been recently proposed, see \citep{Wabersich2019} and references therein, to use projections of the inputs delivered by the RL policy $\vect\pi_{\vect\theta}$ into \textit{safe sets}, which is known by construction to ensure the safety of the system. The construction of the safe set can, e.g., rely on specific knowledge of the system, or robust model predictive control techniques. The projection then operates as a safeguard that prevents RL from taking unsafe decisions, and adopts the safe decision that is the closest to the RL policy when RL is unsafe. 

In this paper, we investigate the interaction between these safe policy projections and the learning process deployed by RL. We show that because the projection modifies the policy developed via RL, it can disrupt the learning process performed such that the learned policy can be suboptimal. The problem occurs both in the context of $Q$-learning and policy gradient approaches using actor-critic methods. We then propose simple techniques to alleviate the problem. In the context of $Q$-learning, we show that the projection technique in general jeopardizes optimality, as it is the projection of a (possibly) optimal policy on a set, and that the problem is best alleviated by relying on a direct minimization of the $Q$ function learned by RL, under the safety constraint that the inputs must belong to the safety set, as proposed in~\citep{Zanon2019b}. In the context of the deterministic policy gradient approaches, we show that, in order to prevent the projections to bias the policy gradient estimations, the actor-critic method must be corrected with a correction which is simple to deploy. In the context of stochastic policy gradient methods, we show that the actor-critic must be constructed in a particular way to prevent the projection from biasing the policy gradient estimations. We finally show that these results extend to the case of a projection performed via robust Model Predictive Control (MPC) techniques.

The paper is structured as follows. Section \ref{sec:Background} provides some background material. Section \ref{sec:Qlearning:Projection} details the projection approach in the context of $Q$-learning, and proposes an approach to address the resulting difficulties. Section \ref{sec:ACMethods} details the projection approach for policy gradient methods, both deterministic and stochastic, and proposes simple actor-critic formulations that prevent the projection from biasing the policy gradient estimations. Section \ref{sec:MPC} extends the results to the case in which the projection is performed via robust MPC. Section \ref{sec:Simulation} proposes a simple simulation example using robust linear MPC in the stochastic policy gradient case, and Section \ref{sec:Conclusion} provides conclusions.

\section{Background} \label{sec:Background}
In the following, we will consider that the dynamics of the real system are possibly stochastic, evolving on continuous state-input spaces. We will furthermore consider stochastic policies $\pi$, taking the form of conditional probability densities ${\pi}\left[\vect{u}\,|\,\vect{x}\right]\,:\, \mathbb R^m \times \mathbb R^n\rightarrow \mathbb R_+,$ denoting the probability density of selecting a given input $\vect u$ when the system is in a given state $\vect x$. We will also consider deterministic policies $\vect\pi\left(\vect x\right) \,:\, \mathbb R^n \rightarrow \mathbb R^m$ delivering $\vect u$ as a function of $\vect x$. 
For a given stage cost $L(\vect x,\vect u) \in\mathbb{R}^n \times \mathbb R^m\rightarrow \mathbb R$ and a discount factor $\gamma \in [0,1]$, the performance of a policy $\pi$ is assessed via the total discounted expected cost
\begin{align}
\label{eq:Return}
J(\pi) = \mathbb{E}_{{ \pi}}\left[\left. \sum_{k=0}^\infty\, \gamma^k L(\vect x_k,\vect u_k)\,\right|\, \vect u_k \sim  \pi\left[\, \cdot\,|\,\vect x_k\right]\, \right],
\end{align}
where $\mathbb{E}_{{ \pi}}$ is the expected value of the closed-loop trajectories under policy $\pi$, including the initial conditions $\vect x_0$.

In the deterministic policy case, the policy in \eqref{eq:Return} takes the form of a Dirac distribution centered at $\vect\pi$. The optimal policy associated to the state transition, the stage cost $L$ and the discount factor $\gamma$ is deterministic and given by
\begin{align}
\label{eq:OptimalPolicy}
\vect\pi_\star =\mathrm{arg} \min_{\vect \pi}\, J(\vect\pi).
\end{align}
Reinforcement Learning seeks to find the parameters $\vect\theta$ such that the parametrized policies $\vect\pi_{\vect\theta}$ or $\pi_{\vect\theta}$ approximate closely $\vect \pi_\star$, using observed state transitions. $Q$-learning methods build the optimal policy approximation indirectly, as the minimizer~\citep{Sutton2018}:
\begin{align}\label{eq:Qlearning:Pi}
{\vect\pi}_{\vect\theta}\left(\vect x\right) = \mathrm{arg}\min_{\vect u}  Q_{\vect\theta}\left(\vect x,\vect u\right),
\end{align}
where $Q_{\vect\theta}$ is an approximation of the true optimal action value function $Q_\star$, solution of the Bellman equations
\citep{Bertsekas2007}:
\begin{subequations}
\begin{align}
V_{\star}\left(\vect x\right) &= \min_{\vect u} Q_{\star}\left(\vect x,\vect u\right), \label{eq:Bellman:Value:0}\\
Q_{\star}\left(\vect x,\vect u\right) &= L(\vect x,\vect u) + \gamma \mathbb{E}\left[V_{\star}(\vect x_{+})\,|\, \vect x,\, \vect u\right]. \label{eq:Bellman:Policy:0} 
\end{align}
\end{subequations}
The approximation $Q_{\vect\theta}\approx Q_\star$ is built using Temporal-Difference or Monte-Carlo techniques.

In contrast, policy gradient techniques manipulate directly the policy parameters according to the policy gradients $\nabla_{\vect\theta}J$~\citep{Sutton1999}. 
Actor-critic techniques evaluate the policy gradient resulting from a stochastic policy as~\citep{Sutton1999}
\begin{align}
\nabla_{\vect\theta}J(\pi_{\vect\theta}) = \mathbb E\left[\nabla_{\vect\theta}\log \pi_{\vect\theta}[\vect u\,|\,\vect x]A_{\pi_{\vect\theta}}\left(\vect x,\vect u\right)\right],
\end{align}
where $A_{\pi_{\vect\theta}}$ is the advantage function associated to the policy $\pi_{\vect\theta}$, defined as
\begin{align}
A_{\pi_{\vect\theta}}\left(\vect x,\vect u\right) &= Q_{\pi_{\vect\theta}}\left(\vect x,\vect u\right) - V_{\pi_{\vect\theta}}\left(\vect x\right), \label{eq:MDP:Afunction:Generic}
\end{align}
and where
\begin{subequations}
\label{eq:Bellman:Policy}
\begin{align}
V_{\pi_{\vect\theta}}\left(\vect x\right) &= \mathbb{E}\left[ L(\vect x,\vect u) + \gamma V_{\pi_{\vect\theta}}(\vect x_{+})\,|\, \vect x,\, \vect u\right], \label{eq:Bellman:Policy:0} \\
Q_{\pi_{\vect\theta}}\left(\vect x,\vect u\right) &= L(\vect x,\vect u) + \gamma \mathbb{E}\left[V_{\pi_{\vect\theta}}(\vect x_{+})\,|\, \vect x,\, \vect u\right],  \label{eq:MDP:Qfunction:Generic}
\end{align}
\end{subequations}
are the value and action-value functions associated to $\pi_{\vect\theta}$.

Similarly, the policy gradient $\nabla_{\vect\theta}J(\vect\pi_{\vect\theta})$ associated to a deterministic policy $\vect\pi_{\vect\theta}$ reads as~\citep{Silver2014}
\begin{align}
\nabla_{\vect\theta}J(\vect \pi_{\vect\theta}) = \mathbb E\left[\nabla_{\vect\theta}\vect \pi_{\vect\theta}(\vect x)\nabla_{\vect u}A_{\vect \pi_{\vect\theta}}\left(\vect x,\vect \pi_{\vect\theta}(\vect x)\right)\right],
\end{align}
where the advantage function $A_{\vect \pi_{\vect\theta}}$ is defined by \eqref{eq:MDP:Afunction:Generic}-\eqref{eq:Bellman:Policy} taken over a Dirac-like policy density corresponding to a deterministic policy. The advantage functions $A_{ \pi_{\vect\theta}}$ and $A_{\vect \pi_{\vect\theta}}$ can be estimated using Temporal-Difference or Monte-Carlo techniques. 

In the context of Reinforcement-Learning, enforcing the safety of the inputs generated by a policy is not trivial~\citep{Garcia2015}. Indeed, for safety-critical systems, discovering unsafe inputs from experiments is overly costly, and is typically rather done in extensive simulation campaigns. As an alternative, recent publications have proposed to approach the safety problem underlying RL by adding a safety layer to the RL process, which serves as a safeguard to the policy, see \citep{Wabersich2019} and references therein. We detail that approach next.

\subsection{Safe Policy}
In this paper, we consider Reinforcement Learning subject to safety limitations. More specifically, we will consider constraints:
\begin{align}
\vect h\left(\vect x,\vect u\right) \leq 0 \label{eq:StageSafety}
\end{align}
that must be respected at all time in order for the system safety to be ensured. Moreover, we will consider a (possibly) state-dependent safe set $\mathbb S\left(\vect x\right)$ such that 
\begin{align}
\vect u_k \in \mathbb S\left(\vect x_k\right),\qquad \forall\, k, \label{eq:OverallSafety0}
\end{align}
entails that \eqref{eq:StageSafety} is satisfied at all times. We ought to stress here the difference between \eqref{eq:StageSafety} and $\mathbb S\left(\vect x\right)$. Satisfying \eqref{eq:StageSafety} at time $k$ entails that the system is safe at that time $k$, while $\mathbb S$ is such that enforcing \eqref{eq:OverallSafety0} at time $k$ entails that the system safety \textit{can} be guaranteed at all time in the future. 
In the following, we will assume that $\mathbb S$ can be described via inequality constraints on $\vect s$, typically different than $\vect h$:
\begin{align}
\mathbb S\left(\vect x\right) = \left\{\, \vect u \,\,\,|\,\,\,  \vect s\left(\vect x,\vect u\right)\leq 0 \right\}. \label{eq:OverallSafety}
\end{align}
Set $\mathbb S\left(\vect x\right)$ can be complex and non-convex. Let us additionally label $\mathbb X$ the set of states $\vect x$ such that $\mathbb S\left(\vect x\right) $ is non-empty, and $\mathbb W = \left\{\, \vect x,\vect u \,\,\,|\,\,\,  \vect s\left(\vect x,\vect u\right)\leq 0 \right\}$. In some applications, the safe set $\mathbb S$ can be computed explicitly using reachability analysis, but that can be prohibitively difficult in general. Inner convex approximations can then be needed. An approach based on an implicit representation has been the object of recent publications \citep{Zanon2019b,Gros2020a}.

Assuming that a safe set $\mathbb S$ is available, a natural approach to ensure the feasibility of a policy ${\vect\pi}_{\vect \theta}$ learned via Reinforcement Learning techniques is to perform a projection into the safe set $\mathbb S$, i.e., to solve online the problem:
\begin{subequations}
\label{eq:SafeRL:Proj}
\begin{align}
\vect\pi_{\vect\theta}^\perp\left(\vect x\right) = \mathrm{arg}\min_{\vect u}&\quad \frac{1}{2}\|\vect u - {\vect\pi}_{\vect\theta}\left(\vect x\right)\|^2\ \\
\mathrm{s.t.}&\quad \vect s\left(\vect x, \vect u\right) \leq 0, \label{eq:SafeRL:Constraint}
\end{align}
\end{subequations}
hence seeking the closest safe input to the RL policy ${\vect\pi}_{\vect\theta}$ under the Euclidian norm $\|\cdot\|$. While \eqref{eq:SafeRL:Proj} imposes safety by construction, the optimality of the projected policy $\vect\pi^\perp_{\vect\theta}$ is, in general, not guaranteed if ${\vect\pi}_{\vect\theta}$ is obtained via RL techniques that disregard the fact that the projection operation \eqref{eq:SafeRL:Proj} takes place. The resulting optimality loss is arguably problem-dependent, and not investigated here. In this paper, we will focus on how \eqref{eq:SafeRL:Proj} can be combined with RL such that optimality of $\vect\pi^\perp_{\vect\theta}$ is achieved. 
\section{Safe $Q$-learning via projection} 
\label{sec:Qlearning:Projection}
In this section we consider the deployment of the $Q$-learning technique under the safety limitation \eqref{eq:OverallSafety}. The minimization in \eqref{eq:Bellman:Value:0} is then restricted to $\mathbb S\left(\vect x\right)$. In the context of $Q$-learning, one seeks to adjust the parameters $\vect\theta$ supporting the function approximation $Q_{\vect\theta}$ such that $Q_{\vect\theta} \approx Q_\star$ is achieved in some sense. The parameters are typically adjusted using Temporal-Difference (TD) or Monte-Carlo techniques, aimed at (approximately) solving the least-squares problem
\begin{align}\label{eq:LSTDQ}
\vect\theta^* = \mathrm{arg}\min_{\vect \theta} \mathbb{E}\left[\left(Q_\star -  Q_{\vect\theta}\right)^2\right].
\end{align}
In a safe-learning context, the expected value in \eqref{eq:LSTDQ} is restricted to the safe state-input set $\mathbb W$, such that $Q_{\vect\theta} \approx Q_\star$ may only hold in $\mathbb W$. The RL policy ${\vect\pi}_{\vect\theta}$ is then selected according to \eqref{eq:Qlearning:Pi}. Let us then investigate the effect of applying the projection \eqref{eq:SafeRL:Proj} on the policy obtained from \eqref{eq:Qlearning:Pi}. To that end, let us introduce a trivial but useful result.

\begin{Lemma} \label{Lemma:Qlearning} Assume that $Q_{\vect\theta}=Q_\star$ holds over $\mathbb W$. Then the optimal policy under the safety requirement \eqref{eq:OverallSafety0}-\eqref{eq:OverallSafety} is provided by:
\begin{subequations}
\label{eq:SafeRL}
\begin{align}
\vect\pi_\star^\mathrm{safe}\left(\vect x\right) = \mathrm{arg}\min_{\vect u}&\quad Q_{\vect\theta}\left(\vect x,\vect u\right)\\
\mathrm{s.t.}&\quad \vect s\left(\vect x, \vect u\right) \leq 0, \label{eq:SafeRL:Constraint:inQ}
\end{align}
\end{subequations}
\end{Lemma}
\begin{pf} By contradiction. Let us assume there is a safe policy $\tilde{\vect\pi}_\mathrm{safe}$ that achieves better closed-loop performance than $\vect\pi_\star^\mathrm{safe}\left(\vect x\right)$ on $\mathbb X$. Because $\tilde{\vect\pi}_\mathrm{safe}$ is safe, it follows that 
\begin{align}
\tilde{\vect\pi}_\mathrm{safe}\left(\vect x\right) \in \mathbb S\left(\vect x\right),\qquad \forall \vect x\in\mathbb X.
\end{align}
If $\tilde{\vect\pi}_\mathrm{safe}$ achieves better closed-loop performance than $\vect\pi_\star^\mathrm{safe}\left(\vect x\right)$, and since $Q_{\vect\theta}=Q_\star$ holds over $\mathbb W$, then there is a $\vect x\in\mathbb X$ such that:
\begin{align}\label{eq:pitilde:optimality}
Q_\star\left(\vect x,\tilde{\vect\pi}_\mathrm{safe}\left(\vect x\right)\right) < Q_\star\left(\vect x,\vect\pi_\star^\mathrm{safe}\left(\vect x\right)\right).
\end{align}
However, since both $\tilde{\vect\pi}_\mathrm{safe}$ and $\vect\pi_\star^\mathrm{safe}\left(\vect x\right)$ are restricted to deliver inputs in $\mathbb S\left(\vect x\right)$, \eqref{eq:pitilde:optimality} is in contradiction with \eqref{eq:SafeRL}. $\hfill\blacksquare$
\end{pf}
\begin{Remark}
Note that because $Q_{\vect\theta}= Q_\star$ may not hold outside of $\mathbb W$, $Q_{\vect\theta}$ may take its minimum outside of $\mathbb S\left(\vect x\right)$ for some states $\vect x\in\mathbb X$. As a result, constraint \eqref{eq:SafeRL:Constraint:inQ} is required in order to generate a safe policy. 
\end{Remark}

\subsection{Projection Approach for $Q$-Learning} \label{sec:QlearningProj}
\begin{figure}
\center
\psfrag{u1}[Bl][Bl][1.1]{$\vect u_1$}
\psfrag{u2}[Bl][Bl][1.1]{$\vect u_2$}
\psfrag{S}[Bl][Bl][1.1]{$\mathbb S(\vect x)$}
\psfrag{piproj}[Br][Br][1]{$\vect\pi_{\vect\theta}^\perp\left(\vect x\right)$}
\psfrag{pistar}[Bl][Bl][1]{$\vect\pi_\star\left(\vect x\right)$}
\psfrag{pi}[Bl][Bl][1]{$\vect\pi_{\vect\theta}\left(\vect x\right)$}
\includegraphics[width=0.275\textwidth,clip]{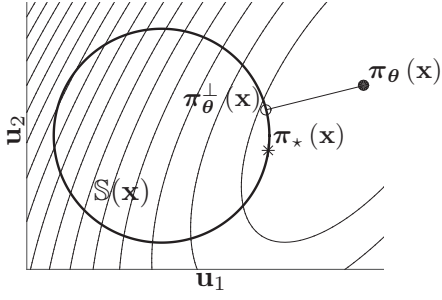}
\caption{Illustration of the possible loss of optimality resulting from using the projection \eqref{eq:SafeRL:Proj} in $Q$-learning.
}
\label{fig:QProblem}
\end{figure}

Consider the projection \eqref{eq:SafeRL:Proj} of the policy \eqref{eq:Qlearning:Pi} obtained via $Q$-learning. We ought to first observe that 
if $Q_{\vect\theta}=Q_\star$ holds over $\mathbb W$, the projected policy is optimal whenever the learned policy ${\vect\pi}_{\vect\theta}\in\mathbb S\left(\vect x\right)$. Unfortunately, this observation does not necessarily extend to the situation where ${\vect\pi}_{\vect\theta}\left(\vect x\right) \notin\mathbb S\left(\vect x\right)$. In order to support this observation, let us consider a trivial example displayed in Fig. \ref{fig:QProblem}. This shows that ${\vect\pi}_{\vect\theta}^\perp\left(\vect x\right) = \vect\pi_\star^\mathrm{safe}\left(\vect x\right)$ does not hold in general. 
However, Lemma \ref{Lemma:Qlearning} readily delivers a way to alleviate this problem: assuming that a $Q$-function approximation $Q_{\vect\theta}\approx Q_\star$ over $\mathbb{Z}$ has been learned, a safe policy can be devised from using $\vect\pi_\star^\mathrm{safe}\left(\vect x\right)$ obtained from \eqref{eq:SafeRL} 
as opposed to a generic projection \eqref{eq:SafeRL:Proj}. 
One then must be careful to include the input restriction $\vect u\in\mathbb S\left(\vect x\right)$ in the evaluation of the TD error underlying the $Q$-learning. When using SARSA, no special care needs to be taken in the learning process, as \eqref{eq:SafeRL} generates all inputs in $\mathbb S\left(\vect x\right)$. An approach to formulate~\eqref{eq:SafeRL} via robust MPC is presented in~\citep{Zanon2019b}.


This section shows that the direct minimization \eqref{eq:SafeRL} of the $Q$ function approximation under the safety constraints is arguably better suited than the two steps approach: \eqref{eq:Qlearning:Pi} followed by \eqref{eq:SafeRL:Proj}. We ought to extend the discussion to the context of the policy gradient methods using actor-critic techniques. This discussion is more technical, and is the object of the next section.

\section{Safe policy gradient via projection} \label{sec:ACMethods}
Policy gradient methods are often preferred over $Q$-learning because they alleviate the known issue that solving the least-squares problem \eqref{eq:LSTDQ} does not necessarily imply that one has found parameter $\vect\theta$ that yields the best closed-loop performance of the policy \eqref{eq:Qlearning:Pi}. Indeed, policy gradient methods seek a direct minimization of the closed-loop cost \eqref{eq:Return} via gradient steps over \eqref{eq:Return}, and therefore yield (at least locally) optimal policy parameters. Similarly to the discussion of Section \ref{sec:Qlearning:Projection}, when deploying policy gradient techniques jointly with a projection on the safe set \eqref{eq:SafeRL:Proj}, the optimality of the resulting policy is unclear. As a matter of fact, we will show in this section that the learning process ought to be corrected in order for the estimation of the gradient of \eqref{eq:Return} to be unbiased. Subsection \ref{sec:DetPiGrad} will cover the deterministic policy gradient case, while subsection \ref{sec:StochPiGrad} will cover the stochastic policy gradient case. 
\subsection{Projected Policy and Deterministic Policy Gradient} \label{sec:DetPiGrad}
In the context of deterministic policies, we will show next that a correction must be applied in the policy gradient computation to account for the safe projection \eqref{eq:SafeRL:Proj}. This correction is provided in the following Proposition.
\begin{Proposition} \label{Prop:DetPolicyGradProjection} 
	Consider the projection \eqref{eq:SafeRL:Proj} where $\|\cdot\|$ stands for the Euclidian norm, and assume that the constraints \eqref{eq:SafeRL:Constraint} satisfy the Linear Constraint Qualification (LICQ) and strict Second-Order Sufficient Conditions (SOSC). The gradient of the projected policy ${\vect\pi}_{\vect\theta}^\perp$ with respect to the policy parameters $\vect\theta$ then reads as:
\begin{subequations}
\begin{align}
\nabla_{\vect\theta}{\vect\pi}_{\vect\theta}^\perp\left(\vect x\right) &= \nabla_{\vect\theta}{\vect\pi}_{\vect\theta}\left(\vect x\right)M\left(\vect x\right),\\ M\left(\vect x\right) &= \mathcal N\left(\mathcal N^\top H\mathcal N\right)^{-1} \mathcal N^\top,
\end{align}
\end{subequations}
where $\mathcal N\in\mathbb R^{m \times n_{\mathbb A}}$ is a state-dependent orthonormal null space to the gradient of the strictly active constraints, i.e.:
\begin{align}
\label{eq:null}
\nabla_{\vect u} \vect s_\mathbb A\left(\vect x,{\vect\pi}_{\vect\theta}^\perp\left(\vect x\right)\right)^\top \mathcal N\left(\vect x\right) = 0,\quad \mathcal N\left(\vect x\right)^\top \mathcal N\left(\vect x\right) = I,
\end{align}
with $\mathbb A$ gathering the set of strictly active constraints $\vect s$, and $H$ is the Hessian associated to \eqref{eq:SafeRL:Proj}.
\end{Proposition}
\begin{pf} 
	The solution to \eqref{eq:SafeRL:Proj} satisfies the KKT conditions:
\begin{align}
\vect r = \matr{c}{\nabla_{\vect u} \mathcal L\left(\vect x, \vect u,\vect \mu\right) \\
\mathrm{diag}(\vect\mu_\mathbb A) \vect s_\mathbb A\left(\vect x, \vect u\right)} &= 0,
\end{align}
where $\mathcal L =\frac{1}{2} \|\vect u - {\vect\pi}_{\vect\theta}\left(\vect x\right)\|^2 + \vect\mu^\top \vect s\left(\vect x, \vect u\right)$. The Implicit Function Theorem guarantees that if LICQ and SOSC hold, the gradient of the projected policy reads as:
\begin{align}\label{eq:SensExpression}
\matr{cc}{H & \nabla_{\vect u} \vect s_\mathbb A \\ \nabla_{\vect u}\vect s_\mathbb A^\top & \bold 0}\matr{c}{\partial \vect u/\partial \vect \theta \\ \partial \vect \mu_\mathbb A/\partial \vect \theta}= -\matr{c}{\nabla_{\vect u\vect\theta} \mathcal L^\top \\ \vect 0}.
\end{align}
We then observe that $\nabla_{\vect u}\vect s_\mathbb A^\top \partial \vect u/\partial \vect \theta = 0$ entails that
\begin{align}
\label{eq:u_Nn}
\nabla_{\vect\theta}{\vect\pi}_{\vect\theta}^\perp\left(\vect x\right)^\top = \frac{\partial \vect u}{\partial \vect \theta} = \mathcal N \vect n,
\end{align}
for some vector $\vect n$. We further observe that:
\begin{align}
\mathcal N^\top\left(H\frac{\partial \vect u}{\partial \vect \theta} + \nabla_{\vect u}\vect s_\mathbb A\frac{\partial \vect \mu_\mathbb A}{\partial \vect \theta} \right) 
&=  -\mathcal N^\top \nabla_{\vect u\vect\theta} \mathcal L^\top \nonumber
\end{align}
follows from \eqref{eq:SensExpression}, such that, using~\eqref{eq:null} and~\eqref{eq:u_Nn} we get
\begin{align}
\mathcal N^\top H\mathcal N \vect n &= - \mathcal N^\top \nabla_{\vect u\vect\theta} \mathcal L^\top.
\end{align}
Since we have
\begin{align}
\nabla_{\vect u\vect\theta} \mathcal L = -\nabla_{\vect\theta}{\vect\pi}_{\vect\theta}, \quad H = \bold I + \nabla^2_{\vect u}\left( \vect\mu_\mathbb A^\top \vect s_\mathbb A\right),
\end{align}
this entails $\vect n = (\mathcal N^\top H\mathcal N)^{-1} \mathcal N^\top \nabla_{\vect\theta}{\vect\pi}_{\vect\theta}^\top$.
\vspace{-0.7cm}
\flushright$\blacksquare$
\end{pf}
Hence the gradient of the projected policy is a form of projection of the gradient of the original policy ${\vect\pi}_{\vect\theta}\left(\vect x\right)$ into the null-space of the safety constraints. We will define $\mathcal N\left(\vect x\right) = I_{m\times m}$ for all $\vect x$ for which all constraints are strictly inactive, and $\mathcal N\left(\vect x\right) = 0_{{m\times 1}}$ for all $\vect x$ where the active constraints fully block the inputs. We observe that the set of states $\vect x$ where some constraints are weakly active---such that the gradient of the policy is only defined in the sense of its sub-gradients---is of zero measure and can therefore be disregarded in the context discussed here. In the particular case of a safety set $\mathbb S$ described as a polytope, such that the constraints $\vect s$ are affine, $H=\bold I$ holds and matrix $M$ simplifies to $M = \mathcal N \mathcal N^\top$.
 
We can then form the Corollary to Proposition \ref{Prop:DetPolicyGradProjection} providing a correct policy gradient evaluation.
\begin{Corollary} Let us assume that \eqref{eq:SafeRL:Proj} fulfills LICQ and SOSC. Then the policy gradient associated to the safe policy ${\vect\pi}^\perp_{\vect\theta}$ reads as:
\begin{align}
\label{eq:truepolicygradient:deterministic}
\nabla_{\vect\theta} J\left({\vect\pi}^\perp_{\vect\theta}\right) &= \mathbb E\left[\nabla_{\vect\theta}{\vect\pi}_{\vect\theta}^\perp\nabla_{\vect u}A_{\vect\pi^\perp_{\vect\theta}}\right] \\
&\quad =\mathbb E\left[\nabla_{\vect\theta}{\vect\pi}_{\vect\theta}M\nabla_{\vect u}A_{\vect\pi^\perp_{\vect\theta}}\right] \nonumber
\end{align}
where $A_{{\vect\pi}^\perp_{\vect\theta}}$ is the advantage function associated to the projected policy ${\vect\pi}^\perp_{\vect\theta}$. All terms in \eqref{eq:truepolicygradient:deterministic} are evaluated at $\vect x,$ $\vect u={\vect\pi}^\perp_{\vect\theta}\left(\vect x\right)$ with $\vect x$ distributed according to the probability density of the states in closed-loop under policy ${\vect\pi}^\perp_{\vect\theta}$.

\end{Corollary}
\begin{pf} 
	We observe that for any $\vect x$ such that no constraint is weakly active, the equality
\begin{align}
\nabla_{\vect\theta}{\vect\pi}_{\vect\theta}^\perp\nabla_{\vect u}A_{{\vect\pi}^\perp_{\vect\theta}}=\nabla_{\vect\theta}{\vect\pi}_{\vect\theta}M\nabla_{\vect u}A_{{\vect\pi}^\perp_{\vect\theta}}\nonumber
\end{align}
holds. If \eqref{eq:SafeRL:Proj} fulfills the LICQ condition, the set of states where some constraints are weakly active is of zero-measure, such that the equality
\begin{align}
\nabla_{\vect\theta} J\left({\vect\pi}^\perp_{\vect\theta}\right) &= \mathbb E\left[\nabla_{\vect\theta}{\vect\pi}_{\vect\theta}^\perp\nabla_{\vect u}A_{{\vect\pi}^\perp_{\vect\theta}}\right] = \mathbb E\left[\nabla_{\vect\theta}{\vect\pi}_{\vect\theta}M\nabla_{\vect u}A_{{\vect\pi}^\perp_{\vect\theta}}\right]\nonumber
\end{align}
holds.
\vspace{-0.5cm}
 \flushright$\blacksquare$
\end{pf}

If deploying the projected policy approach \eqref{eq:SafeRL:Proj} and an actor-critic method not accounting for the projection operation, the policy gradient will generally be such that:
\begin{align}
\label{eq:naivepolicygradient:deterministic}
{\nabla_{\vect\theta} J}\left({\vect\pi}^\perp_{\vect\theta}\right) \neq \mathbb E\left[\nabla_{\vect\theta}{\vect\pi}_{\vect\theta}\nabla_{\vect u}A_{{\vect\pi}^\perp_{\vect\theta}}\right],
\end{align}
where the projection matrix $M\left(\vect x\right)$ is omitted. This omission will, in general, produce a biased policy gradient \eqref{eq:naivepolicygradient:deterministic} if the policy projection is not accounted for in the RL method when computing the policy gradient. It is therefore recommended to form and use the projection matrix $M$ when computing the policy gradient.

It can be advantageous in some cases to adopt a stochastic policy gradient method instead of the deterministic one discussed in this section. In the stochastic policy gradient, the same question arises regarding the learning process being biased by the projection in the safe set. We discuss this case in the next subsection.

\subsection{Projected Policy and Stochastic Policy Gradient} \label{sec:StochPiGrad}

\begin{figure}
\begin{center}
	\psfrag{u1}[Bl][Bl][1.1]{$\vect u_1$}
	\psfrag{u2}[Bl][Bl][1.1]{$\vect u_2$}
	\psfrag{piproj}[Br][Br][1]{$\vect\pi^\perp\left(\vect x\right)$}
	\psfrag{pistar}[Bl][Bl][1]{$\vect\pi_\star\left(\vect x\right)$}
	\psfrag{pi}[Bl][Bl][1]{$\vect\pi_{\vect\theta}\left(\vect x\right)$}
	\includegraphics[width=0.6\linewidth,clip,trim=0 40 0 120]{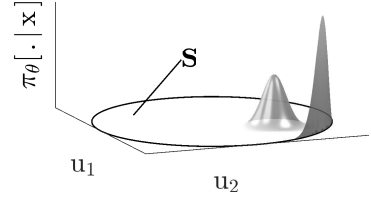}
	\caption{Illustration of the Dirac-like effect resulting from projecting a Normally distributed stochastic policy on a safe set, chosen as a circle here.}
	\label{fig:DiracEffect}
\end{center}
\end{figure}

When using a stochastic policy gradient technique, the inputs are chosen as samples $\vect u_\mathrm{s}$ drawn from a parametrized conditional probability density representing the policy:
\begin{align}
\label{eq:Stoch:Policy}
\vect u_\mathrm{s} \sim \pi_{\vect \theta}[\,\cdot\,|\,\vect x\,].
\end{align}
The safe projection then ought to be performed over the samples $\vect u_\mathrm{s}$, i.e.:
\begin{subequations}
\label{eq:SafeRL:Proj:Stochastic}
\begin{align}
\vect\pi^\perp_{\vect\theta}\left(\vect x,\vect u_\mathrm{s}\right) = \mathrm{arg}\min_{\vect u}&\quad \frac{1}{2}\|\vect u - \vect u_\mathrm{s}\|^2\ \\
\mathrm{s.t.}&\quad \vect s\left(\vect x, \vect u\right) \leq 0. \label{eq:SafeRL:Constraint:Stoch}
\end{align}
\end{subequations}
The inputs generated by $\vect\pi_{\vect\theta}^\perp\left(\vect x,\vect u_\mathrm{s}\right) $ are safe by construction. The resulting projected policy is itself stochastic, as it results from the nonlinear transformation \eqref{eq:SafeRL:Proj:Stochastic} of the probability density $\pi_{\vect\theta}[\,\cdot\,|\,\vect x\,]$. Let us label the probability density resulting from the projection of the stochastic policy $\pi_{\vect \theta}[\,\cdot\,|\,\vect x\,]$ via \eqref{eq:SafeRL:Proj:Stochastic} as $\pi_{\vect\theta}^\perp[\,\cdot\,|\,\vect x\,] $. Unfortunately, since the projection operator defined by \eqref{eq:SafeRL:Proj:Stochastic} is not injective, the density $\pi^\perp_{\vect\theta}[\,\cdot\,|\,\vect x\,] $ can adopt a ``Dirac-like" structure on the boundary $\partial\mathbb S$ of the safe set $\mathbb S$, due to the fact that sets of inputs of dimension larger than one is projected onto a single point on $\partial\mathbb S$. This issue is illustrated in Fig. \ref{fig:DiracEffect}.
As a result, the score function of $\pi^\perp_{\vect\theta}$ is not trivially defined, and the construction of the policy gradient of $\pi^\perp_{\vect\theta}$ is not obvious. The following proposition shows that a trivial modification of the stochastic policy gradient allows one to circumvent this difficulty. 

\begin{Proposition} \label{Prop:StochPolicyGardient} The policy gradient associated to $\pi^\perp_{\vect \theta}$ is given by the actor-critic equation: 
\begin{align}
\label{eq:CoolStuff}
{\nabla_{\vect\theta}J}\left(\pi^\perp_{\vect\theta}\right) = \mathbb E\left[\nabla_{\vect\theta}\log \pi_{\vect \theta}[\vect u_\mathrm{s}|\vect x]A_{\pi^\perp_{\vect\theta}}(\vect x,\vect u)\right],
\end{align}
where $\vect u = \vect\pi_{\vect\theta}^\perp\left(\vect x,\vect u_\mathrm{s}\right)$ is the input obtained from \eqref{eq:SafeRL:Proj:Stochastic} satisfying LICQ and SOSC, and the expected value operator $\mathbb E[\cdot]$ is taken over the state and input distribution obtained in closed-loop under the projected stochastic policy $\pi^\perp_{\vect\theta}[\,\cdot\,|\,\vect x\,]$. 
\end{Proposition}
\begin{pf} 
	In order to build a proof using simple arguments, let us consider the interior-point approximation of the projection problem \eqref{eq:SafeRL:Proj:Stochastic}:
\begin{align}
	\hspace{-6pt}\vect\pi_\tau\hspace{-1pt}\left(\vect x,\vect u_\mathrm{s}\right) \hspace{-1pt}=\hspace{-1pt} \mathrm{arg}\min_{\vect u} \frac{1}{2}\|\vect u - \vect u_\mathrm{s}\|^2 \hspace{-1pt}-\hspace{-1pt} \tau{\sum}_i\log \hspace{-1pt}\left(-\vect h_i\hspace{-1pt}\left(\vect x, \vect u\right)\right), \hspace{-2pt} \label{eq:SafeRL:Constraint:IP}
\end{align}
such that $\left\|\vect\pi_\tau\left(\vect x,\vect u_\mathrm{s}\right) - \vect\pi^\perp_{\vect\theta}\left(\vect x,\vect u_\mathrm{s}\right)\right\| = O(\tau)$ holds. Let us define $\pi^\tau[\,\cdot\,|\,\vect x\,]$ the density resulting from transforming $\pi_{\vect \theta}[\,\cdot\,|\,\vect x\,]$ via \eqref{eq:SafeRL:Constraint:IP}. 
If \eqref{eq:SafeRL:Proj:Stochastic} satisfying LICQ and SOSC, then \eqref{eq:SafeRL:Constraint:IP} is locally bijective in $\vect u_\mathrm{s}$, and the score function associated to $\pi^\tau$ is well-defined. The associated  policy gradient reads as:
\begin{align}
\label{eq:CorrectStochPolGradient:IP}
{\nabla_{\vect\theta}J}\left(\pi^\tau\right) = \mathbb E\left[\nabla_{\vect\theta}\log \pi^\tau[\vect u|\vect x]A_{\pi^\tau}(\vect x,\vect u)\right],
\end{align}
where $\vect u = \vect\pi_\tau\left(\vect x,\vect u_\mathrm{s}\right)$. Let us further define function $\vect \pi_\tau^{-1}$ the local inverse of $\vect\pi^\tau$ at $\vect u_\mathrm{s}$, i.e.,
\begin{align}\label{eq:piinverse}
\vect \pi_\tau^{-1}\left( \vect x, \vect\pi_\tau\left(\vect x,\vect u\right)\right) = \vect u,
\end{align}
holds in a neighborhood of $\vect u_\mathrm{s}$. The existence of \eqref{eq:piinverse} is guaranteed for $\tau >0$ if \eqref{eq:SafeRL:Proj:Stochastic} satisfies LICQ and SOSC. We then observe that the transformation \eqref{eq:SafeRL:Constraint:IP} of the density $\pi_{\vect\theta}$ yields:
\begin{align}
\pi^\tau\left[\vect u\,|\,\vect x\right] &= \pi\left[\,  \vect \pi_\tau^{-1}\left( \vect x, \vect u\right) \,|\,\vect x\,\right] \det\left(\frac{\partial \vect \pi_\tau^{-1}\left( \vect x, \vect u\right)}{\partial \vect u}\right).
\end{align}
For $\vect u_\mathrm s$ given, \eqref{eq:SafeRL:Constraint:IP} is independent of $\vect\theta$, such that
\begin{align}
\nabla_{\vect\theta}\det\left(\frac{\partial \vect \pi_\tau^{-1}\left( \vect x, \vect u\right)}{\partial \vect u}\right) = 0.
\end{align}
As a result, the score function of $\pi^\tau$ reads as:
\begin{align}
\label{eq:CrazyResult}
\nabla_{\vect\theta}\log \pi^\tau\left[\vect u\,|\,\vect x\right] &=\nabla_{\vect\theta}\log  \pi_{\vect \theta}\left[\,  \vect \pi_\tau^{-1}\left( \vect x, \vect u\right) \,|\,\vect x\,\right]\nonumber\\
&=\nabla_{\vect\theta}\log  \pi_{\vect \theta}\left[\,  \vect u_\mathrm{s} \,|\,\vect x\,\right],
\end{align}
where $\vect u_\mathrm{s}$ is the sample corresponding to $\vect u$ obtained from \eqref{eq:SafeRL:Constraint:IP}. Combining \eqref{eq:CrazyResult} and \eqref{eq:CorrectStochPolGradient:IP}, we observe that
\begin{align}
{\nabla_{\vect\theta}J}\left(\pi^\tau\right) = \mathbb E\left[\nabla_{\vect\theta}\log  \pi_{\vect \theta}\left[\,  \vect u_\mathrm{s} \,|\,\vect x\,\right] A_{\pi^\tau}(\vect x,\vect\pi_\tau\left(\vect x,\vect u_\mathrm{s}\right))\right].\nonumber
\end{align}
The equivalence between \eqref{eq:SafeRL:Constraint:IP} and \eqref{eq:SafeRL:Proj:Stochastic} for $\tau\rightarrow 0$ implies that \eqref{eq:CoolStuff} holds. 
\vspace{-0.5cm}
\flushright$\blacksquare$
\end{pf}
\begin{Remark} Proposition \ref{Prop:StochPolicyGardient} allows one to use the projection technique in the context of RL based on a stochastic policy approach together with an actor-critic technique, where the score function of the unprojected policy can be used in conjunction with the advantage function associated to the projected policy. The score function of the unprojected policy must then be evaluated on the unprojected sample, rather than on the projected input applied to the system.
\end{Remark}
As mentioned earlier, the construction of the constraints $\vect s$ underlying the safe set $\mathbb S\left(\vect x\right)$ can be difficult. We extend next the proposed results to MPC-based techniques allowing one to build the safety constraints implicitly, via model predictive control techniques.
\section{MPC-based projections} \label{sec:MPC}
It is in general difficult to build the safe set $\mathbb S$ from condition \eqref{eq:StageSafety}. Indeed, an input $\vect u$ applied at a given time $k$ can  have lasting consequences and, while not endangering the system at time $k$, jeopardize its safety in the future. In order to alleviate this problem, the safety constraints can be built implicitly via Model Predictive Control (MPC) techniques. 
In that context, let us consider
\begin{align}
\label{eq:Dispersion}
\vect X_k\left(\vect x,\vect u_{0},\vect\pi^\mathbb S\right),\quad k=0,\ldots,\infty
\end{align}  
an outer approximation of the trajectory dispersion of the real system starting from the initial conditions $\vect x$, hence $\vect X_0\left(\vect x,\vect u_{0},\vect\pi^\mathbb S\right) = \vect x$, and subject to the input sequence $\vect u_{0},\,\vect u_k = \vect\pi^\mathbb S\left(\vect x_k\right)$, where $\vect\pi^\mathbb S$ is an arbitrary policy. The safe set can then be described as an inner approximation:
\begin{align}
\mathbb S\left(\vect x\right)  &\subseteq \left\{\,\vect u_0 \,\,\, | \,\,\, \exists\, \vect \pi^{\mathbb S}\quad  \mathrm{s.t.} \quad  \vect h\left(\vect x_k ,\vect \pi^\mathbb S(\vect x_k)\right)\leq 0\right . \nonumber\\&\hspace{5em}\left. \forall\,\vect x_k\in\vect X_k\left(\vect x,\vect u_{0},\vect\pi^\mathbb S\right),\quad \forall\, k>0 \right\},
\end{align}
which can then be used in \eqref{eq:SafeRL:Proj} or \eqref{eq:SafeRL:Proj:Stochastic}. If MPC techniques are used, a generalization of the projection technique can be considered. In the deterministic policy case, one can then use the generic robust formulation:
\begin{subequations}
\label{eq:SafeRL:Proj:Deterministic:MPC}
\begin{align}
\hspace{-4pt}\left(\vect u_{0},\vect\pi^{\mathbb S}\right) \hspace{-2pt}\left(\vect x_0\right)  \hspace{-1pt}= \arg\min_{ \vect u_{0},\vect\pi^{\mathbb S} }&\, \frac{1}{2} \|\vect u_0 \hspace{-1pt}-\hspace{-1pt} \vect \pi_{\vect \theta}\hspace{-1pt}\left(\vect x_0\right) \hspace{-1pt}  \|^2  \hspace{-1pt}+\hspace{-1pt} \phi\left(\vect\pi^{\mathbb S},\vect \pi_{\vect \theta}\right)\hspace{-0pt}\\
 \mathrm{s.t.}&\quad  \vect h\left(\vect x ,\vect u_0\right)\leq 0, \label{eq:SafeRL:Constraint:0} \\
 &\quad \vect h\left(\vect x_k ,\vect \pi^\mathbb S(\vect x_k)\right)\leq 0,\label{eq:SafeRL:Constraint:1}\\&\quad \forall\,\vect x_k\in\vect X_k\left(\vect x,\vect u_{0},\vect\pi^\mathbb S\right),\,\, \forall\, k>0.\nonumber
\end{align}
\end{subequations}
We can then select $\vect\pi_{\vect\theta}^\perp\left(\vect x\right) = \vect u_{0}\left(\vect x\right) $ as a safe control input. In the stochastic policy case, the equivalent formulation reads as: 
\begin{subequations}
\label{eq:SafeRL:Proj:Stochastic:MPC}
\begin{align}
\left(\vect u_{0},\vect\pi^{\mathbb S}\right)\hspace{-2pt}\left(\vect x_0\right)  = \arg\min_{ \vect u_{0},\vect\pi^{\mathbb S} }&\ \frac{1}{2}\|\vect u_0 - \vect u_\mathrm s \|^2  + \phi\left(\vect\pi^{\mathbb S},\pi_{\vect \theta}\right)\\
 \mathrm{s.t.}&\  \eqref{eq:SafeRL:Constraint:0}- \eqref{eq:SafeRL:Constraint:1},
\end{align}
\end{subequations}
where $\vect u_\mathrm s\sim \pi_{\vect\theta}[\,\cdot\,|\,\vect x\,]$ is a sample drawn from the stochastic policy. The cost function $\phi$ in \eqref{eq:SafeRL:Proj:Deterministic:MPC} can be independent of $\vect\pi_{\vect\theta}$, or, e.g., any metric in the functional space underlying the deterministic policies $\vect \pi_{\vect \theta}$ and $\vect \pi^{\mathbb S}$. A similar construction can be done for $\phi$ in \eqref{eq:SafeRL:Proj:Stochastic:MPC}.

The following corollaries show that Propositions \ref{Prop:DetPolicyGradProjection} and \ref{Prop:StochPolicyGardient} hold in the context of \eqref{eq:SafeRL:Proj:Deterministic:MPC} and  \eqref{eq:SafeRL:Proj:Stochastic:MPC} under some conditions.

\begin{Corollary} Proposition \ref{Prop:DetPolicyGradProjection} holds for \eqref{eq:SafeRL:Proj:Deterministic:MPC} with:
\begin{subequations}
\label{eq:Prop1:proj}
\begin{align}
\nabla_{\vect u\vect\theta} \mathcal L &= -\nabla_{\vect\theta}{\vect\pi}_{\vect\theta} + \nabla_{\vect u\vect\theta}\rho_{\vect \theta}\left(\vect u_0,\vect x\right), \\ H &= \bold I + \nabla^2_{\vect u_0} \rho_{\vect \theta}\left(\vect u_0,\vect x\right)+ \nabla^2_{\vect u_0}\left( \vect\mu_\mathbb A^\top \vect s_\mathbb A\right).
\end{align} 
\end{subequations}
where 
\begin{subequations}
\label{eq:rho}
\begin{align}
\rho_{\vect \theta}\left(\vect u_0,\vect x\right)= \min_{\vect\pi^{\mathbb S}} &\quad \phi\left(\vect\pi^{\mathbb S},\vect \pi_{\vect \theta}\right) \quad \mathrm{s.t.}\quad \eqref{eq:SafeRL:Constraint:1},
\end{align}
\end{subequations}
\end{Corollary}
\begin{pf}  Problem \eqref{eq:SafeRL:Proj:Deterministic:MPC} can be put in the form:
\begin{subequations}
\label{eq:SafeRL:Proj:Deterministic:MPC:split}
\begin{align}
 \vect u_{0} =  \mathrm{arg}\min_{ \vect u_{0} }&\ \ \frac{1}{2}\|\vect u_0 - \vect \pi_{\vect \theta}\left(\vect x\right) \|^2  + \rho_{\vect \theta}\left(\vect u_0,\vect x\right)\\
 \mathrm{s.t.}&\ \ \vect s\left(\vect x ,\vect u_0\right)\leq 0, \label{eq:SafeRL:Constraint:0:split} 
\end{align}
\end{subequations}
One can then readily observe that Proposition \ref{Prop:DetPolicyGradProjection} applies to \eqref{eq:SafeRL:Proj:Deterministic:MPC:split}, with \eqref{eq:Prop1:proj}.
\vspace{-0.5cm}
\flushright$\blacksquare$
\end{pf}

\begin{Corollary} \label{Cor:MPCStochastic} 
	The results of Proposition \ref{Prop:StochPolicyGardient} hold for \eqref{eq:SafeRL:Proj:Stochastic:MPC} \textcolor{black}{if function $\phi$ is independent of $\vect\theta$.}
\end{Corollary}
\begin{pf} 
	Problem \eqref{eq:SafeRL:Proj:Stochastic:MPC} can be put in the form:
\begin{subequations}
\label{eq:SafeRL:Proj:Stochastic:MPC:split}
\begin{align}
 \vect u_{0} =  \mathrm{arg}\min_{ \vect u_{0} }&\ \ \frac{1}{2}\|\vect u_0 - \vect u_\mathrm s \|^2  + \rho_{\vect \theta}\left(\vect u_0,\vect x\right)\\
 \mathrm{s.t.}&\ \  \vect s\left(\vect x ,\vect u_0\right)\leq 0, \label{eq:SafeRL:Constraint:0:split} 
\end{align}
\end{subequations}
where $\rho_{\vect\theta}$ is based on $\phi\left(\vect\pi^{\mathbb S},\pi_{\vect \theta}\right)$. One can verify that Proposition \ref{Prop:StochPolicyGardient} is independent of the choice of cost function in the projection as long as it is independent of $\vect \theta$, and holds as long as it satisfies LICQ/SOSC. As a result, if $\phi$ is independent of $\vect\theta$, Proposition \ref{Prop:StochPolicyGardient} readily applies to \eqref{eq:SafeRL:Proj:Stochastic:MPC:split}.
\vspace{-0.25cm}
\flushright$\blacksquare$
\end{pf}
If function $\phi$ depends on $\vect\theta$, more elaborate techniques must be used, see \citep{Gros2020a}.

\section{Simulated Example} \label{sec:Simulation}
\begin{figure}
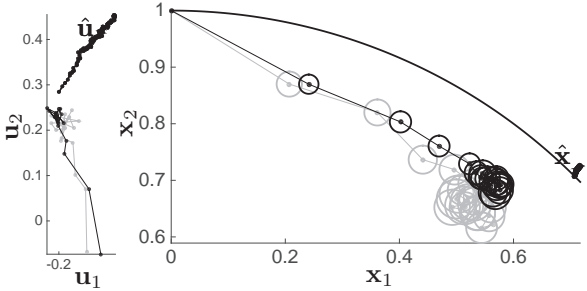

\center
\psfrag{u1}[Bl][Bl][1]{$\vect u_1$}
\psfrag{u2}[Bl][Bl][1]{$\vect u_2$}
\psfrag{x1}[Bl][Bl][1]{$\vect x_1$}
\psfrag{x2}[Bl][Bl][1]{$\vect x_2$}
\psfrag{xref}[Bl][Bl][1]{$\hat{\vect x}$}
\psfrag{uref}[Br][Br][1]{$\hat{\vect u}$}
\includegraphics[width=0.08\textwidth,clip]{Figures/Traju.eps}\includegraphics[width=0.35\textwidth,clip]{Figures/Trajx.eps}
\caption{Illustration of the input and state trajectories at the beginning (grey curves) and end (black curves) of the learning. The circles in the right graph display the state standard deviation. The markers show the evolution of the MPC references $\hat{\vect u}_\mathrm{ref},\,\hat{\vect x}_\mathrm{ref}$.
}
\label{fig:traj}
\end{figure}

\begin{figure}
\center
\psfrag{s}[Tc][Tc][.8]{RL batch $k$}
\psfrag{theta}[Bc][Bc][1]{$\vect\theta_k$}
\psfrag{k1}[Bl][Bl][1]{$K_{11}$}
\psfrag{k2}[Bl][Bl][1]{$K_{21}$}
\psfrag{k3}[Bl][Bl][1]{$K_{12}$}
\psfrag{k4}[Bl][Bl][1]{$K_{22}$}
\includegraphics[width=0.35\textwidth,clip]{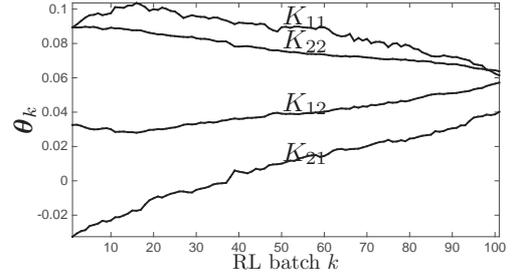}
\caption{Evaluation of the policy parameters $\vect\theta$ associated to the feedback matrix $K$ in \eqref{eq:RL:Policy} over the learning
}
\label{fig:parameters}
\end{figure}

\begin{figure}
\center
\psfrag{s}[Tc][Tc][.8]{RL batch $k$}
\psfrag{J}[Bc][Bc][1.0]{$J(\pi_{\vect\theta_k})/J(\pi_{\vect\theta_0})$}
\includegraphics[width=0.35\textwidth,clip]{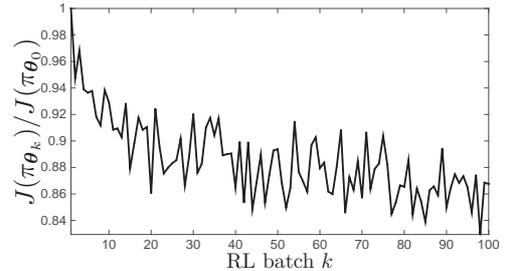}
\caption{Evaluation of the policy performance \eqref{eq:Return} over the learning, normed to 1.
}
\label{fig:cost}
\end{figure}

In this section, we present a simple example illustrating Corollary \ref{Cor:MPCStochastic}. Let us consider the dynamic system:
\begin{align}
\label{eq:System}
\vect x_{k+1} = \matr{cc}{\phantom{-}\cos a& \sin a \\ -\sin a & \cos a} \vect x_k + \vect u_k + \vect n_k,
\end{align}
where $a= 20^\circ$, $\vect n_k\in\mathbb R^2$ is truncated Normal centred of covariance $\Sigma_{\vect n} = 0.1\mathrm I$, and restricted to a ball of radius $0.1$, i.e., $\vect n_k \in \mathcal B\left(0,0.1\right)$. We consider a safety constraint:
\begin{align}\label{eq:safe:example}
\vect h\left(\vect x\right) = \vect x^\top \vect x - 1 \leq 0.
\end{align}
We will use the baseline cost:
\begin{align}
L\left(\vect x,\vect u\right) = 10^{-2} \|\vect x-\vect x_{\mathrm {ref}}\|^2 + \|\vect u-\vect u_{\mathrm {ref}}\|^2.
\end{align}
The MPC will be based on the noise-free model
\begin{align}
\label{eq:Model}
\bar{\vect x}_{k+1} = 1.1\matr{cc}{\phantom{-}\cos \hat a& \sin \hat a \\ -\sin \hat a & \cos \hat a} \bar{\vect x}_k + \vect u_k,
\end{align}
where $\hat a = 25^\circ$, and the policy $\vect \pi^\mathbb S$ will be selected as:
\begin{align}
\vect \pi^\mathbb S\left(\vect x,\vect u\right) = {\vect u} - K^\mathbb S\left(\vect x - \bar{ \vect x}_k\right),
\end{align}
where $K^\mathbb S$ is the LQR corresponding to \eqref{eq:Model} for $Q,R = \mathrm I$. We can represent the dispersion set as a ball, i.e.,
\begin{align}
\label{eq:Dispersion}
\vect X_k\left(\vect x,\vect u_{0},\vect\pi^\mathbb S\right) = \mathcal B\left(\bar{\vect x}_k,r_k\right)
\end{align}  
of radius 
$r_{k+1} = \|A\|_\infty r_k + \max_{\vect n \in \mathcal B\left(0,0.1\right)} \|\vect n\|$, and $r_0=0$. We then build the robust MPC scheme:
\begin{subequations}
\label{eq:RobustMPC:Example}
\begin{align}
\hspace{-4pt}\vect u_{0,\ldots,N-1}(\vect x) = \arg \hspace{-4pt} \min_{\hspace{-12pt}\vect u_{0,\ldots, N-1}}&\, \frac{1}{2}\|\vect u_0 - \vect u_\mathrm s \|^2 \hspace{-1pt} + \hspace{-1pt}\sum_{k=1}^{N-1} \gamma^kL\left(\bar{\vect x}_{k},\vect u_k \right)\\
\mathrm{s.t.} & \ \eqref{eq:Model},\,\,\,\, \bar{\vect x}_0 = \vect x,\,\, \,\,\vect h\left({\vect x}_{k}\right) \leq 0 \\
&\quad \forall {\vect x}_{k}\in\mathcal B\left(\bar{\vect x}_k,r_k\right),\quad \forall k.
\end{align}
\end{subequations}
with $\gamma = 0.9$, and use $\vect\pi^\perp_{\vect\theta}\left(\vect x\right) = \vect u_{0}\left(\vect x\right)$. We consider the stochastic policy $\pi_{\vect \theta}$ delivering the samples $\vect u_\mathrm s$ as Normal, centred at $\bar{\vect\pi}_{\vect \theta}\left(\vect x\right)$, and of isotropic covariance $\sigma_{\pi}\mathrm I$, 
where
\begin{align}
\label{eq:RL:Policy}
\bar{\vect\pi}_{\vect \theta}\left(\vect x\right) = \hat{\vect u}_\mathrm{ref} - K\left(\vect x - \hat{\vect x}_\mathrm{ref}\right),
\end{align}
and the policy parameters are $\vect\theta = \left\{\hat{\vect u}_\mathrm{ref},\,\hat{\vect x}_\mathrm{ref},\, K\right\}$. A batch approach was used to compute the policy gradients, using \eqref{eq:CoolStuff}, using 30 batches of duration 20, and LSTDV/LSTDQ techniques. The initial condition $\vect x_0 = \matr{cc}{0&1}^\top$ was used. The MPC horizon is $N=10$. A linear compatible advantage function approximator was used, built upon a quadratic value function approximation. Fig. \ref{fig:parameters} displays the evolution of the policy parameters through the learning, Fig. \ref{fig:cost} shows the evaluation of the closed-loop performance, and Fig. \ref{fig:traj} shows the evolution of the system trajectories through the learning process.

\section{Conclusion} \label{sec:Conclusion}
In this paper, we discussed the projection approach as a method to enforce the safety of a policy learned via RL. We showed that the approach is detrimental in the context of $Q$-learning, and that a direct minimization of the $Q$ function under the safety constraints is arguably more suited. We showed that in the context for deterministic policies, the actor-critic method needs a simple correction in order for the policy gradient estimation to be unbiased. Similarly, in the context of stochastic policies, the actor-critic needs to be constructed in a very specific way in order for the policy gradient estimations to be unbiased. We showed that the results extend to the case of a projection performed via Robust MPC.
\bibliography{syscop}

\end{document}